\documentclass[%
reprint,
superscriptaddress,
preprintnumbers,
nofootinbib,
 amsmath,amssymb,
 aps,
prd,
floatfix,
twocolumn,
showpacs,
superscriptaddress,
]{revtex4-2}
\usepackage[normalem]{ulem}
\usepackage{slashed}
\usepackage{graphicx}
\usepackage{dcolumn}
\usepackage{bm}
\usepackage[dvipsnames]{xcolor}    
\definecolor{lcolor}{rgb}{0.5,0,0}
\definecolor{citcolor}{rgb}{0,0,1}
\usepackage[breaklinks,colorlinks,urlcolor=blue,citecolor=citcolor,linkcolor=lcolor,linktoc=all]{hyperref}
\usepackage[T1]{fontenc}
\usepackage[utf8]{inputenc}
\usepackage{orcidlink}
\usepackage{footnote}
\usepackage{bbold}
\usepackage{caption}

\usepackage{soul}

\renewcommand{\d}{\mathrm d}
\newcommand{\Hr}{\mathcal{H}}
\newcommand{\OGW}{\Omega_\textsc{\rm GW}}
\newcommand{\rGW}{\rho_\textsc{\rm GW}}

\newcommand\bea{\begin{equation}}
\newcommand\eea{\end{equation}}

\usepackage{verbatim}

\begin{document}

\title{Signatures from pion condensation and lepton flavor asymmetries in the cosmological gravitational wave background}


\author{Osvaldo {\sc Ferreira}\orcidlink{0000-0002-6711-8308}}
 \affiliation{
 Instituto de F\'isica, Universidade de S\~ao Paulo, R. do Mat\~ao, 1371, S\~ao Paulo, Brazil, 05508-090
}
 \affiliation{
 Instituto de F\'\i sica, Universidade Federal do Rio de Janeiro,
 CEP 21941-972 Rio de Janeiro, RJ, Brazil 
}

\author{Eduardo S. {\sc Fraga}\orcidlink{0000-0001-5340-156X}} 

\affiliation{
 Instituto de F\'\i sica, Universidade Federal do Rio de Janeiro,
 CEP 21941-972 Rio de Janeiro, RJ, Brazil 
}

\author{Jürgen {\sc Schaffner-Bielich}\orcidlink{0000-0002-0079-6841}}
\affiliation{Institut für Theoretische Physik, J. W. Goethe Universität,
 Max von Laue-Str. 1, 60438 Frankfurt am Main, Germany}


\date{\today}
\begin{abstract}

Large lepton flavor asymmetries at the QCD epoch could generate a pion condensation phase in the early Universe. For large enough tau lepton flavor asymmetries, the speed of sound can exceed the conformal value, leaving a distinctive imprint on the low-frequency gravitational wave (GW) spectrum from causal sources. Beyond probing the formation of a pion condensation phase, the detection or non-detection of this signature in future Pulsar Timing Arrays (PTAs) datasets would provide a novel constraint on lepton asymmetries in the early Universe. We determine the corresponding GW spectrum and its tilt, comparing them with the standard case of vanishing lepton asymmetry. Finally, we discuss the implications for the stochastic GW background reported by PTAs, using the NANOGrav 15-year dataset.
\end{abstract}

\maketitle



\emph{Introduction.}
Understanding the quantum chromodynamics (QCD) epoch in the early Universe is crucial for describing the formation of visible matter. It is at this time that quarks bonded to form hadrons and acquired most of their mass through chiral symmetry breaking. This now corresponds to almost all the mass of ordinary matter in the Universe. Probing the QCD epoch is, however, a difficult endeavor since only indirect information can be inferred from the Cosmic Microwave Background (CMB) or Big Bang Nucleosynthesis (BBN), which refer to later times in the cosmic evolution. The observation of the stochastic background of gravitational waves (SBGW) may be the only way to directly probe the first second of cosmic evolution. In particular, phenomena occurring at the QCD era, at temperatures $T\sim 150$ MeV, would impact SBGW in the nanohertz range, where Pulsar Timing Arrays (PTAs) are the most effective probes. In fact, PTA collaborations have reported evidence for a nanohertz gravitational wave background (NANOGrav~\cite{NANOGrav:2023gor, NANOGrav:2023hde}, European PTA (EPTA)~\cite{EPTA:2023sfo, EPTA:2023fyk}, Parkes~\cite{Reardon:2023gzh} and the Chinese PTA \cite{Xu:2023wog}).

Nevertheless, current SGWB observations lack the precision to distinguish whether the origin of the signal is of cosmological or astrophysical origin \cite{Figueroa:2023zhu, NANOGrav:2023gor, NANOGrav:2023hvm, Franciolini:2023wjm}. Moreover, the prediction of cosmological SGWBs is usually dependent on the microscopic nature of the sources. Refs.~\cite{Hook:2020phx, Figueroa:2019paj, Caprini:2009fx, Cai:2019cdl} have made the important observation that, for a number of causal sources (e.g, first-order phase transions~\cite{Caprini:2009fx, Cai:2019cdl, Athron:2023xlk}, some scenarios of inflation ~\cite{Kolb:1990ey, Sa:2007pc}, cosmic strings~\cite{Ghoshal:2025iil, Datta:2025vyu}, etc.), the low frequency or (IR) tail of the gravitational wave spectrum exhibits a universal behavior $\OGW \sim f^3$ in a radiation dominated Universe. The important implication of this observation is that equations of state with deviations from pure radiation ($w=\frac{1}{3}$), can leave imprints on this IR tail, also called Causality Tail (CT), rendering them a powerful tool to probe the microphysics of the early Universe.

\begin{figure}[t]
    \centering
    \includegraphics[width=0.8\linewidth]{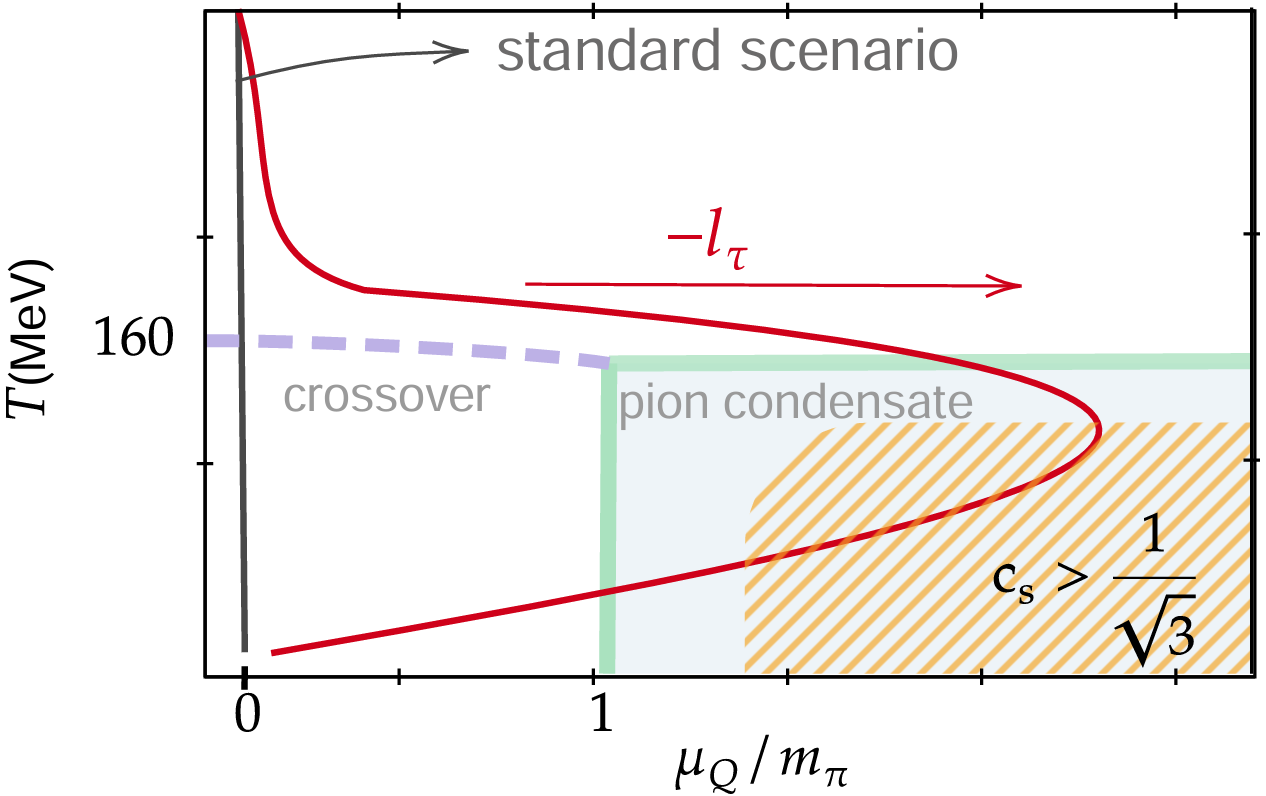}
    \caption{A schematic cartoon showcasing how the cosmic trajectory could enter a region with a speed of sound surpassing the conformal value. The QCD phase diagram shown in the background is adapted from Ref.\cite{Brandt:2022hwy} and is based on Lattice QCD simulations.}
    \label{fig: Cartoon_PD_speed_of_sound_above_CF.png}
\end{figure}

Recently, Ref. \cite{Franciolini:2023wjm} has evaluated what would be the expected imprints on the CT from the equation of state predicted by Lattice QCD. When using Lattice QCD at zero charge and baryon chemical potential, this work and others \cite{Saikawa:2018rcs, Brzeminski:2022haa, hajkarim2019a, Hajkarim:2019csy} rely on the standard assumption that the lepton asymmetries in the early Universe were of the same order as the baryon asymmetry, $l \sim b \sim 10^{-11}$. 
However, if large lepton asymmetries were present at the QCD epoch, charge neutrality and the conservation of baryon and lepton numbers, $B$ and $L$, would lead to high charge chemical potentials $\mu_Q$~\cite{Vovchenko2020crk, Middeldorf-Wygas:2020glx}. When $\mu_Q \approx m_\pi$, with $m_\pi$ being the pion mass, pions can form a Bose-Einstein condensate~\cite{Brandt:2017oyy,Brandt:2022hwy, Brandt:2018bwq}. In fact, Refs.~\cite{Vovchenko2020crk,Middeldorf-Wygas:2020glx,Ferreira:2025zeu} have shown that this phase could be reached in the early Universe for high enough lepton asymmetries, see Fig.~\ref{fig: Cartoon_PD_speed_of_sound_above_CF.png}.

Constraining primordial lepton asymmetries and distinguishing lepton-asymmetric scenarios from the standard scenario is challenging ~\cite{Domcke:2025lzg,Domcke:2025jiy, Froustey:2021azz, Froustey:2024mgf}. Frequently, a first-order phase transition is suggested as a consequence of non-vanishing lepton asymmetries, which could lead to observable signals by generating primordial gravitational waves~\cite{schwarz2009, gaoFunctional2022, Zheng:2024tib, Ferreira:2025zeu}. This would be a major finding, but previous work suggests that the baryon chemical potentials are not high enough to generate such transitions~\cite{DiClemente:2025awt,Middeldorf-Wygas:2020glx, Wygas2018otj, Formaggio:2025nde}. An alternative would be hitting a first-order line when entering the pion condensation boundary~\cite{Ferreira:2025zeu}, but the existence of this line still needs to be confirmed by Lattice QCD~\cite{Brandt:2017oyy}.
Another possibility discussed in Ref.~\cite{Vovchenko2020crk, Bodeker:2020stj}
is that such asymmetries could produce shifts of standard observables such as the primordial GW spectrum from inflation~\cite{Vovchenko2020crk} or the primordial black hole spectrum~\cite{Bodeker:2020stj}. However, a similar effect in the GW spectrum may be generated by different parameters in inflationary models, which reduces its power as a distinguishing feature~\cite{Guzzetti:2016mkm}.
 
Therefore, although these effects are certainly relevant and their combination could still support lepton flavor asymmetric scenarios, it is crucial to look for distinct features that could disentangle them from the standard scenario and accurately probe QCD phenomena in the early Universe.

In this paper we address this gap and report on a characteristic imprint that the formation of pion condensation could leave on the low-frequency tail of the SGWB of causal sources. More precisely, given that large enough lepton asymmetries were present at the QCD epoch, the formation of a pion condensate may have made the equation of state parameter $w(T)$ of the Universe peak~\cite{Brandt:2024dle, Brandt:2022hwy, Abbott:2023coj, Abbott:2024vhj} above the conformal value ($w=\frac{1}{3}$), producing a distinctive signature in the low-frequency part of the GW spectrum of causal sources. Moreover, the derivative of the spectrum with respect to frequency, the so called GW spectral tilt, has been shown to be directly related to $w(T)$~\cite{Hook:2020phx} and is specially sensitive to this effect.

Our main result is establishing a connection between three different observations: first, the CT is independent of the microphysics of the source, but is sensitive to the equation of state of the Universe~\cite{Hook:2020phx}. Second, high lepton asymmetries can lead to the formation of pion condensation in the early Universe~\cite{Vovchenko2020crk, Middeldorf-Wygas:2020glx}. And third, it is a well-established prediction from Lattice QCD that the speed of sound can peak at high enough values of charge (or isospin) chemical potentials inside the pion condensation phase~\cite{Brandt:2024dle, Brandt:2022hwy, Abbott:2023coj, Abbott:2024vhj}. We therefore propose a way in which future GW observations could confirm or constrain the possible formation of pion condensation in the early Universe and the associated presence of lepton asymmetries in this epoch.



\emph{High-isospin QCD equation of state and the peak in the speed of sound.}
To determine the equation of state of the Universe and its corresponding cosmic trajectories, we need to find for a given temperature, the lepton, baryon and charge chemical potentials of the cosmic plasma, while imposing charge neutrality and the conservation of $B$ and $L$ (for details see Refs.~\cite{schwarz2009, Wygas2018otj, Middeldorf-Wygas:2020glx, Vovchenko2020crk, Ferreira:2025zeu, Formaggio:2025nde}). In the simplest case, if we assume that the primordial lepton asymmetries $l_\alpha = (n_\alpha + n_{\nu_{\alpha}})/s$ ($\alpha=e, \mu, \tau$) were of the same order as the baryon asymmetry $b = n_B/s$ at the time of BBN,  with $s$ the total entropy density, all chemical potentials vanish, and the EoS of the Universe at the QCD epoch can be determined from Lattice QCD~\cite{Borsanyi:2016ksw, Saikawa:2018rcs}. However, it has been pointed out that flavor-asymmetric lepton asymmetries can shift the trajectory to higher values of $\mu_Q$ and $\mu_B$~\cite{Middeldorf-Wygas:2020glx, Wygas2018otj, Vovchenko2020crk, Ferreira:2025zeu}. The modeling of such scenarios is, however, more challenging since Lattice QCD cannot probe the full QCD phase diagram at finite $\mu_Q$ and $\mu_B$.

Nevertheless, first-principle methods can still make clear predictions about high-isospin QCD, which we may take as a guide. Pion condensation occurs when $\mu_Q$ (= $2 \mu_I$) exceeds the value of $\approx 140$ MeV for temperatures $\lesssim 160$ MeV \cite{Brandt:2017oyy}. At zero temperature, Refs. \cite{Abbott:2024vhj, Abbott:2023coj} have reported that, at high isospin, the QCD speed of sound has a peak above the conformal limit at an isospin chemical potential $\mu_I \sim 1.5 m_\pi$ and this is consistent with chiral perturbation theory and perturbative QCD\footnote{It is important to notice that Ref. \cite{Abbott:2024vhj} has used a pion mass of $m_\pi = 170$ MeV.}. At finite temperature (but $T < 145$ MeV), Ref. \cite{Brandt:2022hwy} has also found\footnote{Notice that Ref. \cite{Brandt:2022hwy} adopts a a different convention for $\mu_I$ than Ref. \cite{Abbott:2024vhj} and the one used here.} that the speed of sound seems to suddenly increase at $\mu_I/(2m_\pi)$. The effect of the temperature seems to be to shift the peak of the speed of sound to higher values of $\mu_I$. As discussed in Refs. \cite{Brandt:2025tkg, Brandt:2019hel, Cuteri:2021hiq} this peak may also be related to the BEC-BCS transition.


\begin{figure}[b]
    \centering
    \includegraphics[width=\linewidth]{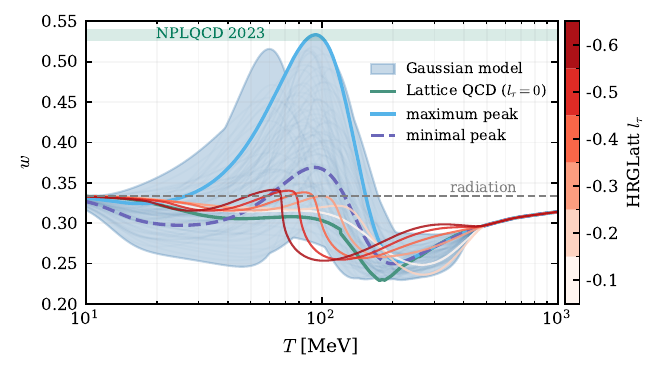}
    \caption{EoS parameter $w$ for different equations of state. The blue band are sampled curves of the gaussian model parametrization and the blue solid and purple dashed lines are examples to illustrate the effect. The green line correspond to the Lattice EoS of Refs.~\cite{Borsanyi:2016ksw, Saikawa:2018rcs, Franciolini:2023wjm} with $l=l_{\tau} =0$, while the green NPLQCD 2023 band refers to a reference value for the peak of $w$ from Lattice QCD at high isospin from Ref.~\cite{Abbott:2023coj}. The red curves correspond to the HRGLatt model for different lepton asymmetries.}
    \label{fig: w models}
\end{figure}

Since there is no first-principle EoS for the QCD epoch in lepton asymmetric scenarios, we build a simple parameterization for $w(T)$ in terms of Gaussians (see the Supplemental Material~\cite{supp} for details) in the temperature range around the QCD epoch. At high and low temperatures, the lepton asymmetric scenarios should approach the standard case~\cite{DiClemente:2025awt, Formaggio:2025nde, schwarz2009, Wygas2018otj} (see Fig.~\ref{fig: Cartoon_PD_speed_of_sound_above_CF.png}). Additionally, we impose the constraint that the peak in $w$ should not exceed the value obtained by the Lattice simulations of Ref.~\cite{Abbott:2023coj} for zero temperature high-isospin QCD and set the QCD crossover dip from Ref.~\cite{Borsanyi:2016ksw} as a lower bound. After sampling $w(T)$ for different parameters controlling the positions, heights and widths of the high-isospin peak and the QCD crossover dip, we show the remaining constrained samples in Fig.~\ref{fig: w models} as a blue band. From these we chose two illustrative curves: the first (blue solid line) reaches the maximum possible peak, and the second (purple dashed line) has a peak which is high enough ($w=0.37$) to produce a peak in the GW tilt that can reach the radiation line (see below).

Moreover, we also show results for the model that we call HRGLatt. It consists of an interpolation of the EoS parameter $w(T)$ determined using the HRG~\cite{Vovchenko2020crk}\footnote{This equation of state can be determined using the \hyperlink{https://thermal-fist.vovchenko.net/}{\texttt{Thermal FIST}}~\cite{Vovchenko:2019pjl}.} for the range\footnote{Notice that Ref.~\cite{Vovchenko2020crk} has shown that one must have $|l_e+l_\mu |\geq 0.1$ to reach a pion condensation phase by setting $l_\tau = - (l_e + l_\tau)$. By construction, the range of (total) lepton asymmetries considered is within the allowed values from BBN and the CMB~\cite{Domcke:2025jiy, Domcke:2025lzg, Froustey:2024mgf, Burns:2022hkq, Akita:2025txo, Akita:2025zvq}.} $l_\tau=[-0.6,-0.1]$ and the Lattice QCD EoS for vanishing lepton flavor asymmetries discussed in Refs.~\cite{Borsanyi:2016ksw, Saikawa:2018rcs} at higher temperatures. The HRG EoS has been used in previous evaluations of cosmic trajectories at high $\mu_Q$~\cite{Vovchenko2020crk, DiClemente:2025awt} and offers a reasonable description of the pion condensation phase, although it likely underestimates the height of the peak, as discussed in Refs.~\cite{Brandt:2025tkg, Ayala:2024sqm, Lopes:2025rvn}.


\emph{Causality tail and tilt in the GW spectrum.} The Causality Tail of the GW spectrum of short-lived sources corresponds to the low-frequency part of the spectrum in which waves have a period and wavelength much larger than the source's temporal and spatial correlations \cite{Hook:2020phx, Cai:2019cdl}. However, this tail is sensitive to the equation of state. 
Therefore, our working hypothesis from now on is that a GW signal may have been generated by a short-lived source at a time $\tau_\star$ (and temperature $T_{\star}$) before the QCD epoch, and we discuss how a peak from high-isospin QCD may affect it. For instance, Ref.~\cite{Franciolini:2023wjm} argues that such a signal must be generated at $T_{\star} \gtrsim 300$ MeV to ensure that, in the frequency ranges discussed here, the signal actually refers to a CT. 
The GW spectrum can be defined in terms of the frequency as 
\begin{equation}
    \OGW (f) \equiv \frac{1}{\rho_\text{c}} \frac{\d \rho_{\rm GW}(f)}{\d \ln f} \, ,
\end{equation}
where $\rho_{\rm GW}$ is the GWs energy density and $\rho_c = 3 H^2/(8 \pi G)$ is the critical energy density, with $H$ the Hubble rate and $G$ is Newton's constant. 

To understand how we can probe the EoS from the CT and its spectrum, let us assume that $w'(\tau) \ll \Hr$, then one can show that \cite{Hook:2020phx, Brzeminski:2022haa} $\Omega_{\rm GW} \sim k^{\frac{1+15 w(\tau)}{1+3 w(\tau)}} \, .$ Where, for each momentum $k$, $w(\tau)$ should be taken at the time $\tau$ when $k$ re-enters the horizon, $k=\Hr(\tau)$. Moreover, one can define the GW spectral tilt $\alpha_{\mathrm{\rm GW}}\equiv \frac{\d \Omega_{\rm GW}}{\d \ln k}$\footnote{Notice that, if one use the definition of Refs. \cite{Hook:2020phx, Brzeminski:2022haa} for the GW spectrum, then the tilt is a second derivative with respect to $\ln k$.}.
Which under this approximation leads to~\cite{Hook:2020phx} $\alpha_{\rm GW} = \frac{1+15 w(\tau)}{1+3 w(\tau)}$. Therefore, one can extract $w(\tau)$ by measuring the tilt of the low-frequency tail of the GW spectrum. Although these equations can be useful for a rough estimate, they may overestimate the effects of rapidly varying EoSs.

\begin{figure}[t]
    \centering
    \includegraphics[width=\linewidth]{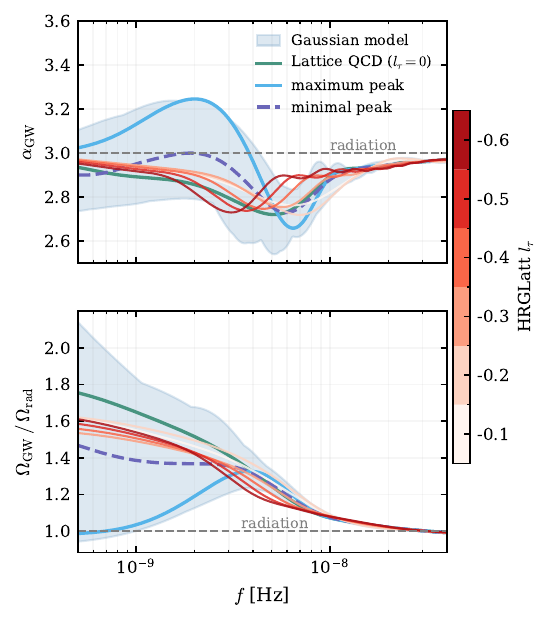}
    \caption{GW tilt (top) and spectrum (bottom) for the different EoSs discussed in the text. We show how an EoSs representing high isospin QCD (and large lepton asymmetries) deviate from the standard prediction of Lattice QCD at vanishing lepton asymmetries.}
    \label{fig: GW tilt and spectrum for different models}
\end{figure}


\emph{GW tilt induced by a non-conformal speed of sound.} Since both $\Omega_{ \rm GW, CT}$ and $\alpha_{\rm GW}$ are directly related to $w(\tau)$, they will both change according to its features. We determine the spectrum by numerically solving the equations of motion for the GW modes $h_k$ for a wide range of momenta $k < k_{\star}$ (see supplemental Material~\cite{supp} for details). We also take $f_{\text{yr}} = 32$ nHz as a reference value for the frequency, as in Ref.~\cite{Franciolini:2023wjm}, so that $\OGW(f_{\text{yr}})=1$. The spectrum is then obtained up to an amplitude $A_{\mathrm{\rm GW}}$, resulting in the plots of Fig.~\ref{fig: GW tilt and spectrum for different models}. In the upper panel of Fig. \ref{fig: GW tilt and spectrum for different models}, we plot the slope of the GW spectrum $\alpha_{\rm GW}$ as a function of frequency for the equations of state shown in Fig.~\ref{fig: w models}. An EoS parameter peaking at $w=0.37$ (purple dashed line) above the conformal value can make $\alpha_{\rm GW}$ have a peak touching the radiation line. Also, if $w(T)$ peaks at the reference value from zero temperature high-isospin QCD from Ref.~\cite{Abbott:2023coj}, this would translate into $\alpha_{\rm GW}$ peaking significantly above the radiation line at $\alpha_{\rm GW}=3.24$ (blue solid line). As for the HRGLatt model, although the predicted peaks in $w(T)$ are not large enough to translate into a distinctive peak in the GW tilt, we can see some relevant qualitative differences, such as the curves shifting upward, getting closer to the radiation line and the QCD crossover dip being shifted to lower temperatures for higher lepton asymmetries.

In the lower panel of Fig.~\ref{fig: GW tilt and spectrum for different models} we plot the GW spectral density $\OGW$ normalized by $\Omega_{\mathrm{rad}}=\left({f}/{f_{\text{yr}}}\right)^3$ as a function of frequency for the different equations of state. We also include the line for the $\sim f^3$ scaling expected for a relativistic free gas (radiation). We see that all EoSs deviate from the pure radiation case for frequencies $\lesssim 10^{-8}$ Hz. The Lattice QCD EoS shows a small bending towards the radiation line at lower frequencies, probably because after the crossover $w(T)$ approaches $1/3$ at lower temperatures. This trend is amplified in the lepton asymmetric cases. Moreover, this bending starts at different points for each case, which seems to be related to the position of the crossover dip and is more pronounced for wider and higher peaks, as illustrated by both the HRGLatt and Gaussian model benchmark cases. This is consistent with Ref.~\cite{Brzeminski:2022haa}, which shows that, in the case of a constant $w(T)$, with $w>\frac{1}{3}$, the CT line should be below the pure radiation line. Finally, notice that, in the HRGLatt cases, even the EoSs that do not peak above the conformal value predict effects on the spectrum.

\begin{figure}[t]
    \centering
    \includegraphics[width=\linewidth]{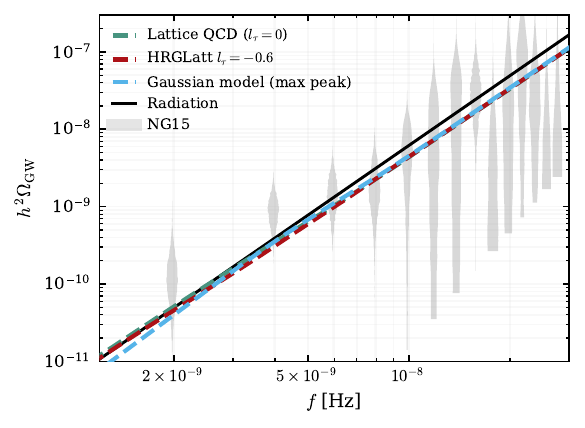}
    \caption{GW CT spectra from Lattice QCD (blue), HRGLatt (red) and Gaussian (orange) EoSs. We use the Maximum a Posteriori (MAP) amplitudes of each model obtained from the NANOGRAV 15 dataset~\cite{NANOGrav:2023gor} (grey violins), see Supplemental material~\cite{supp}. The MAPs were obtained using the \href{https://andrea-mitridate.github.io/PTArcade/}{\texttt{PTArcade}} \cite{Mitridate:2023oar, andrea_mitridate_2023, lamb2023need} wrapper.}
    \label{fig: NG15 fit}
\end{figure}

As we discussed previously, GWs in the nanohertz range can be probed by PTAs, and a number of PTA collaborations have reported evidence for a SGWB~\cite{NANOGrav:2023gor, NANOGrav:2023hde, EPTA:2023sfo, EPTA:2023fyk, Reardon:2023gzh, Xu:2023wog}. However, the current datasets lack the precision to make conclusive statements about the effects proposed in this work. Nevertheless, it is natural to consider the implications of our results in terms of the current observations and we do it here using the NANOGRAV 15 years data~\cite{NANOGrav:2023gor} due to its smaller uncertainties compared to the other datasets. The results are shown in Fig.~\ref{fig: NG15 fit}, where we perform a Bayesian analysis to find the best fit for the logarithmic amplitude $\log_{10}A$ of the GW spectrum predicted by the HRGLatt model with $l_{\tau}=-0.6$, the maximum-peak case of the Gaussian model, and the Lattice EoS ($l_\tau =0$) which is discussed in detail in Ref.~\cite{Franciolini:2023wjm}. The results were obtained using the \href{https://andrea-mitridate.github.io/PTArcade/}{\texttt{PTArcade}} wrapper, with the \texttt{ceffyl}~\cite{lamb2023need} mode and Hellings-Downs correlations. More details can be found in the Supplemental Material~\cite{supp}. The most interesting qualitative feature is that the maximum-peak EoSs does not cross the radiation line at any point, in contrast to the other EoSs considered.


\emph{Constraining lepton asymmetries.} In Ref.~\cite{Vovchenko2020crk} it is shown that the relevant quantity to generate values of $\mu_Q$ that are high enough to form a pion condensate is $|l_{\tau}|$ (assuming $l_e + l_{\mu} + l_{\tau} = 0$). This can be simply understood as follows. As discussed in Ref.~\cite{DiClemente:2025awt}, at lower temperatures and roughly until the maximum $\mu_Q$ is reached, $\tau$ particles are Boltzmann suppressed. Therefore, in this regime we may neglect $\mu_\tau$, and $l_\tau$ is dominated by the tau neutrino asymmetry, i.e, $l_{\tau} \approx {n_{\nu_{\tau}}}/{s(T)}$ .  Moreover, conservation equations imply~\cite{Wygas2018otj,Ferreira:2025zeu} $\mu_\alpha =\mu_{L_\alpha}-\mu_Q$ and $\mu_{\nu_\alpha}=\mu_{L_\alpha}$, with $\alpha=e, \mu, \tau$. So, if we neglect $\mu_\tau$ these relations imply that $\mu_Q = \mu_{\nu_{\tau}}$, and we can show that $\mu_Q^3 \approx 6\pi^2 l_{\tau} s(T)$, with $\mu_Q > T$, which is compatible with the values found in calculations of cosmic trajectories~\cite{Vovchenko2020crk, Ferreira:2025zeu}. Therefore, the specific temperature dependence of $\mu_Q$ comes from $s(T)$ while $l_\tau$ sets how large $\mu_Q$ values can be, which dictates the possible appearance of a peak in $w(T)$.

Notice that the relation of $|l_{\tau}|$ and $\mu_Q$ is very robust since it is rooted in the conservation equations, and this implies that the values for $\mu_Q$ carry only a weak model dependence~\cite{DiClemente:2025awt}. However, the behavior of the speed of sound and $w(T)$ is model dependent. Therefore, a more robust prediction for the spectrum and its implications requires improving Lattice QCD at finite temperature and high isospin (or $\mu_Q$). This, combined with improved data from PTAs, can confirm or rule out the signatures proposed earlier in this work. This would not only provide us information on QCD phenomenology, e.g., pion condensation and the behavior of the speed of sound, but would also supply valuable information about primordial lepton asymmetries, which are very hard to constrain before BBN~\cite{Domcke:2025lzg, Froustey:2021azz, Froustey:2024mgf}.

\emph{A short stiff-dominated period.} An interesting interpretation for the short period in which the equation of state could be larger than the conformal value is that it would correspond to a stiff-dominated period. Stiffer phases in the early Universe have been proposed in the context of inflationary cosmology~\cite{Figueroa:2019paj, Giovannini:1998bp, Boyle:2007zx}. It is intriguing that something analogous may occur within the Standard Model, in lepton-asymmetric scenarios. This implies that, for a short period, high-lepton asymmetries may make the energy density dilute faster during the QCD epoch.


\emph{Conclusions.}
We showed that, if a cosmological GW signal was generated before the QCD epoch by a short-lived source, the observation of its low-frequency spectrum by PTAs could probe the possible formation of pion condensation in the early Universe, while also constraining primordial lepton asymmetries. 
Because the speed of sound of high-isospin QCD has a distinct peak above the conformal limit, this can generate a very characteristic imprint.
To our knowledge, there is no similar effect stemming from any other phenomena in the Standard Model, making it a very distinctive signature. Although we cannot rule out a Beyond Standard model effect that could mimic such signature, it would need to be fine tuned to only affect the QCD sector in this specific way. From a QCD perspective, since the only way known to reach the required values of charge chemical potentials is through large lepton asymmetries, this effect can also be used as a proxy for the presence of lepton asymmetries in the early Universe.
Finally, our discussion is in a range of frequencies where PTAs are most sensitive~\cite{Sesana:2025udx, Kelley:2025yud}. With a better signal-to-noise ratio from PTA observations, and improved Lattice QCD calculations at finite temperature and isospin density (along the lines of \cite{Brandt:2024dle}), one may use the (non)detection of this effect as a way to constrain the possible lepton asymmetries at the QCD epoch from SGWB observations.

\begin{acknowledgments}
We thank Dietrich Bodeker for useful comments and Volodymyr Vovchenko for providing tables and instructions about the use of Thermal FIST. OF and ESF acknowledge the support by INCT-FNA (Process No. 464898/2014-5), CAPES (Finance Code 001), CNPq, and FAPERJ. OF was also partially supported by the São Paulo Research Foundation (FAPESP). JSB acknowledges support by the Deutsche Forschungsgemeinschaft (DFG, German Research Foundation) through the CRC-TR 211 'Strong-interaction matter under extreme conditions'– project number 315477589 – TRR 211.
\end{acknowledgments}

\newpage

\vspace{10cm}


\bibliographystyle{apsrev4-1}
\bibliography{references.bib}

\section*{Supplemental Material}

\section{Causality Tail}

As shown in Ref.~\cite{Hook:2020phx}, if we are interested only in the Causality Tail part of the spectrum, the equation of motion for the modes can be simplified as follows. Since the modes contributing to the CT have a period much longer than the source's duration, the source behaves essentially as a Dirac delta, $J(\tau,k)\to J_\star \delta(\tau-\tau_\star)$. Also, an initially vanishing GW mode gets a velocity $h'_k=J_\star$, with the prime denoting a derivative with respect to conformal time $\tau$, shortly after $\tau_\star$, and starts to evolve according to~\cite{Hook:2020phx, Brzeminski:2022haa, Franciolini:2023wjm}
\begin{equation}\label{eq: CT GW modes eom}
\begin{gathered}
h''_k(\tau) + 2 \mathcal H(\tau) h'_k(\tau) + k^2 h_k(\tau) =0 \,,\\
h_k(\tau_\star)=0\,, \quad h'_k(\tau_\star)=J_\star \, ,
\end{gathered}
\end{equation}
where the conformal Hubble rate $\Hr(\tau)$ and $a(\tau)$ must be found by solving the Friedmann equations and are therefore dependent on $w$.
The energy density of GWs can be written as 
\begin{equation}\label{eq: GW propagation}
\rGW(\mathbf x,\tau) = \sum_{r,s=+,\times} 
  \frac{1}{32 \pi G a^2} \left \langle 
  h^{\prime\, r}_{ij}(\mathbf x,\tau) 
  h^{\prime\, s}_{ij}(\mathbf x,\tau) \right\rangle \,
\end{equation}
for the relevant modes that have entered the horizon at late times and started oscillating. Then, we can use $\OGW (f) = \frac{1}{\rho_\text{c}} \frac{\d \rho_{\rm GW}(f)}{\d \ln f} \,$ to obtain the GW spectrum.

\section{Models for the equation of state}\label{appendix: EoS models}

\subsection{Gaussian model}

In this section, we describe the Gaussian model used in the discussion. The essential idea is to generate a number of $w(T)$ curves that could mimic the qualitative behavior expected for the speed of sound from Lattice QCD \cite{Brandt:2017oyy, Brandt:2022hwy, Brandt:2024dle,Abbott:2023coj, Abbott:2024vhj} and from Refs. \cite{Vovchenko2020crk, Ferreira:2025zeu} that investigated the possible realization of pion condensation in the early Universe. 

Each function $w(T) = P/\varepsilon$ is modeled as a superposition of generalized Gaussians, sometimes called super-Gaussians, taking the conformal value $w = 1/3$ as a baseline. The model smoothly interpolates between a low-temperature limit ($w(T) \rightarrow \frac{1}{3}$) and Lattice QCD data at high temperatures. We define 
\begin{equation}\label{eq: w gaussian model}
    w(T) = \frac{1}{3} + G_\mathrm{low} + G_\mathrm{peak} - G_\mathrm{core} - G_\mathrm{tail},
\end{equation}
where $G_\mathrm{peak}$ produces the high-isospin peak, $G_\mathrm{low}$ and $G_\mathrm{tail}$ control the low-$T$ and high-$T$ behavior, respectively, and $G_\mathrm{core}$ is associated to the crossover dip. Each of these Gaussian components are given by functions of the form
\begin{equation}
    G_i(T) = A_i \exp\!\left(-\left(\frac{|T - T_i|}{\sigma_i}\right)^{n_i}\right),
\end{equation}
where $A_i$ is the amplitude, $T_i$ is the center temperature, $\sigma_i$ is the width, and $n_i$ is the shape exponent. Notice that for $n_i = 2$ this reduces to a standard Gaussian.
Moreover, below $T \approx 30$ MeV we use a sigmoid function $S(T;\, T_c, \Delta) = \frac{1}{1 + e^{-(T - T_c)/\Delta}}$ to smoothly approach the conformal value. At high temperatures, $T_{c,\mathrm{high}} = 400$ MeV, the model transitions to a cubic spline interpolation of Lattice QCD data using the function $w(T) = (1 - S)\, w_\mathrm{Gaussian} + S\, w_\mathrm{lattice}$.
The parameters of the four components in Eq.~\eqref{eq: w gaussian model} are chosen to mimic the presence of a crossover (the dip) and a high-isospin peak. To scan the parameter space for the possible $w(T)$, we allow the eight parameters listed in Table~\ref{tab: ranges} to vary and used Latin-hypercube sampling\cite{McKay1979LHS, Virtanen2020SciPy}. A a curve is kept only if it satisfies the constraints: the peak must not exceed the NPLQCD upper bound, the dip must not fall below the Lattice QCD minimum, and $w$ must reach the pure radiation value ($1/3$) below $10$ MeV and the lattice value over $400$--$1000$~MeV. We then randomly vary the parameters until $100$ candidates for $w(T)$, satisfying the constraints, are selected. The remaining parameters are held fixed at the values in Table~\ref{tab: base parameters}.

\begin{table}[h]
\centering
\caption{Sampled parameter ranges.}
\label{tab: ranges}
\begin{tabular}{@{}llcl@{}}
\toprule
Parameter & Meaning & Range & Units \\
\hline
$A_{\rm peak}$   & isospin peak height            & $[0.05,\ 0.32]$ & --- \\
$T_{\rm peak}$   & peak position                  & $[60,\ 110]$    & MeV \\
$\sigma_{\rm peak}$ & peak width                  & $[25,\ 80]$     & MeV \\
$A_{\rm core}$   & crossover dip depth            & $[0.0,\ 0.13]$  & --- \\
$T_{\rm core}$   & dip position                   & $[85,\ 200]$    & MeV \\
$\sigma_{\rm core}$ & dip width                   & $[80,\ 250]$    & MeV \\
$A_{\rm low}$    & low-$T$ shoulder height        & $[0.0,\ 0.05]$  & --- \\
$A_{\rm tail}$   & high-$T$ deficit amplitude     & $[0.03,\ 0.08]$ & --- \\
\hline
\end{tabular}
\end{table}

\begin{table}[htb]
\centering
\caption{Fixed parameters, held constant during the sampling of the ranges in Table~\ref{tab: ranges}.}
\label{tab: base parameters}
\begin{tabular}{@{}lccc@{}}
\toprule
Component & $T$ [MeV] & $\sigma$ [MeV] & exponent $n$ \\
\hline
low      & 10  & 20  & 1 \\
peak     & --- & --- & 2 \\
core     & --- & --- & 2 \\
tail     & 320 & 350 & 2 \\
\hline
\end{tabular}
\end{table}

\subsection{HRGLatt equation of state}

The second set of models used in the paper combines through a two‑point cubic Hermite interpolant~\cite{deBoor2001Splines, NumericalRecipes2007} a Hadron Resonance Gas (HRG) equation of state, computed with the Thermal-FIST~\cite{Vovchenko:2019pjl} package, with Lattice QCD data at high temperatures. We explored the range $l_\tau =-0.1,...,-0.6$ for the lepton asymmetries. Moreover, the bridge between the two Thermal-FIST EoSs and the Lattice EoS starts at $\min(190,\,T^{\rm FIST}_{\max}-10)$ MeV and ends at $500$ MeV.

\begin{figure}[b]
    \centering
    \includegraphics[width=\linewidth]{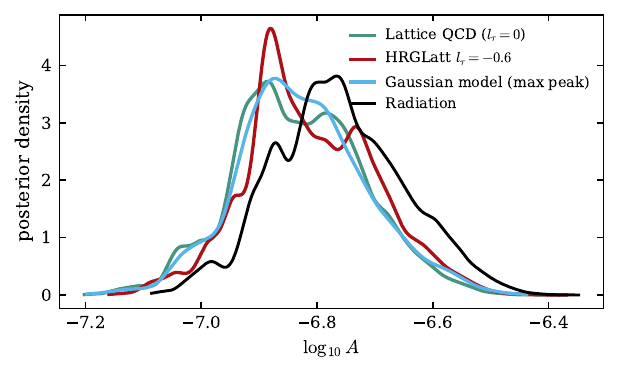}
    \caption{Posterior distributions of the logarithmic amplitude $\log_{10}A_{\mathrm{GW}}$ for the the equation-of-state models discussed in the main text. The pure-radiation ($\sim f^3$) case was also included for reference. All curves are fit to the same Hellings-Downs-correlated NG15 free spectrum.}
    \label{fig: posterior NANOGRAV Fit}
\end{figure}

\section{Bayesian fit to NANOGRAV 15 years data}\label{appendix: posterior values}

As discussed in the main text, the GW spectrum $\OGW$ is obtained up to an amplitude $A$. We use the NANOGRAV 15 years dataset to find the maximum a posteriori (MAP) amplitudes for each model. 
We use the \href{https://andrea-mitridate.github.io/PTArcade/}{\texttt{PTArcade}} wrapper with the \texttt{ceffyl}~\cite{lamb2023need} mode, fitting the Hellings--Downs-correlated NG15 free spectrum ($14$ frequency bins, $2\times10^{6}$ MCMC samples) with a log-uniform prior for $A_{\mathrm{GW}}$ over the $\left[-10,-4\right]$ interval. The resulting MAP values are $\log_{10}A_{\mathrm{Latt}} = -6.88$, $\log_{10}A_{\mathrm{HRGLatt}} = -6.88$ and $\log_{10}A_{\mathrm{gauss}} = -6.87$ (with $68\%$ credible intervals all close to $[-6.93,\,-6.70]$). For reference, we also performed a fit for a pure-radiation ($\sim f^3$) spectrum to the same data, which gives $\log_{10}A_{\mathrm{rad}} = -6.76$. All posteriors are shown in Fig.~\ref{fig: posterior NANOGRAV Fit}.

\end{document}